\documentclass[a4paper,11pt]{article}
\pdfoutput=1
\usepackage{jcappub}
\usepackage[T1]{fontenc}
\usepackage{url}
\usepackage{hyperref}
\usepackage{mathtools}
\usepackage{tensor}
\usepackage{palatino}
\usepackage{tikz}
\usepackage{multirow}
\usepackage{tabularx}
\usetikzlibrary{shapes.geometric, arrows}

\title{Late-time acceleration and inflation in a Poincar\'e gauge cosmological model}

\author[a,b]{Hongchao Zhang}
\author[a]{Lixin Xu}
\affiliation[a]{Institute of Theoretical Physics, School of Physics, \\ Dalian University of Technology, Dalian, 116024, P. R. China}
\affiliation[b]{Institute for Gravitation and the Cosmos, \\ Penn State University, University Park, PA 16801, U.S.A.}
\emailAdd{zhanghc@mail.dlut.edu.cn}
\emailAdd{lxxu@dlut.edu.cn}
\date{\today}

\abstract{A self-consistent model which can unify Starobinsky inflation and the $\Lambda$CDM model in the framework of Poincar\'e gauge cosmology (PGC) is studied in this work, without extra inflaton and ``dark energy''. We start from the general nine-parameter PGC Lagrangian and get two Friedmann-like analytical solutions with the certain ghost- and tachyon-free conditions on the Lagrangian parameters. The scalar torsion $h$-determined solution is consistent with Starobinsky cosmology in the early-time and the pseudo-scalar torsion $f$-determined solution contains naturally a constant ``dark energy'' density, $3(\alpha-1)^2/8B_0$, which covers the $\Lambda$CDM model in the late-time.
According to the latest observations, we estimate the magnitudes of parameters: $\alpha-1\simeq 8.07\times10^{-56}$, $B_0\simeq5.76\times10^{-28}GeV^{-2}$ in the natural units.}

\begin{document}
\maketitle
\flushbottom

\section{Introduction}
Inflation and late-time acceleration are two significant accelerated expansion periods of the Universe in the standard model (SM) framework of cosmology based on Einstein's general relativity (GR) and observations.
Latest observations \cite{aghanim2018planck} indicate good consistency with the standard spatially-flat $6$-parameter $\Lambda$CDM cosmology having a power-law spectrum of adiabatic scalar perturbations, from polarization, temperature, and lensing, separately and in combination.
The joint constraint with baryon acoustic oscillation (BAO) measurements on spatial curvature is consistent with a flat universe, $\Omega_K=0.001\pm0.002$. Also combining with Type Ia supernovae (SNe), the equation of state (EOS) parameter of ``dark energy'' is measured to be $w_0=-1.03\pm0.03$, which is consistent with a cosmological constant.
However, although the $\Lambda$CDM cosmology can accurately describe the evolution of the universe from a phenomenological perspective, the value of vacuum energy density estimated from quantum field theory is $10^{121}$ times larger than the observed value \cite{copeland2006dynamics}, and there is still no evidence of the existence of ``dark energy''.
On the other hand, the constraints on inflation \cite{akrami2018planck} support the key prediction of the standard single-field (inflaton) inflation models.
However, any single-field inflation model faces the origin of the scalar field.
As a single-field model, Starobinsky inflation given by a Lagrangian $\tilde{R}+\tilde{R}^2/6M^2$ plus some small non-local terms (which are crucial for reheating after inflation) is an internally self-consistent cosmological model, which possess a (quasi-)de Sitter stage in the early Universe with slow-roll decay, and a graceful exit to the subsequent radiation-dominated Friedmann-Lema\^{i}tre-Robertson-Walker (FLRW) stage \cite{starobinsky1980new,vilenkin1985classical,mijic1986r}.
This is one of the most appealing from both theoretical and observational perspectives among different models of inflation \cite{castellanos2018higher}.
The motivation is natural to unify both Starobinsky inflation and the $\Lambda$CDM in one model, as people tried in \cite{nojiri2003modified,nojiri2007unifying,cognola2008class,nojiri2008modified,elizalde2011nonsingular,nojiri2011unified,elizalde2017beyond,odintsov2017unification}, from a perspective of $f(R)$ gravity.

Besides adding directly higher-order curvature invariants to the Hilbert-Einstein (HE) action, another more fundamental way to generalize GR from the geometric and gauge perspectives was introduced systematically since 1970's \cite{hehl1976general,blagojevic2013gauge}, which is called the Poincar\'e gauge gravity (PGG).
As the maximum group of Minkowski spacetime isometrics, the Poincar\'e group is the semidirect product of the translation group and the rotation group, which has $10$ degrees of freedom in total.
If constructing a gauge field theory based on the local invariance of the Poincar\'e group, gravity will be represented by two sets of independent gauge fields: the canonical $1$-forms $\theta$s (the dual co-vectors of tetrads $e$s) and the spin-connections $\omega$s, corresponding to the translations and the rotations, respectively.
Analogous to the Yang-Mills theory, one can verify that torsion $T$ and curvature $R$ are just their gauge field strengths.
According to Noether's theorems, the symmetries of translations and rotations lead to two sets of conserved objects: the energy-momentums and the spin-angular momentums.
Furthermore, the energy-momentum can be connected through Einstein's equation with curvature, and the spin-angular momentum with torsion through Cartan's equation, which mean that the sources of spacetime curvature and torsion are energy-momentum and spin of matter, respectively.
The above is the basic idea of PGG, which follows the schemes of the Yang-Mills theory.
From the geometrical perspective, the spacetime extends from Riemann's to Riemann-Cartan's, where curvature measures the difference of a vector after parallel transporting along an infinitesimal loop, and torsion measures the failure of closure of the parallelogram made of the infinitesimal displacement.
To show the extension of PGG to GR, we plot the following diagram:
\begin{figure}[htbp]
	\centering
	\includegraphics[scale=0.6]{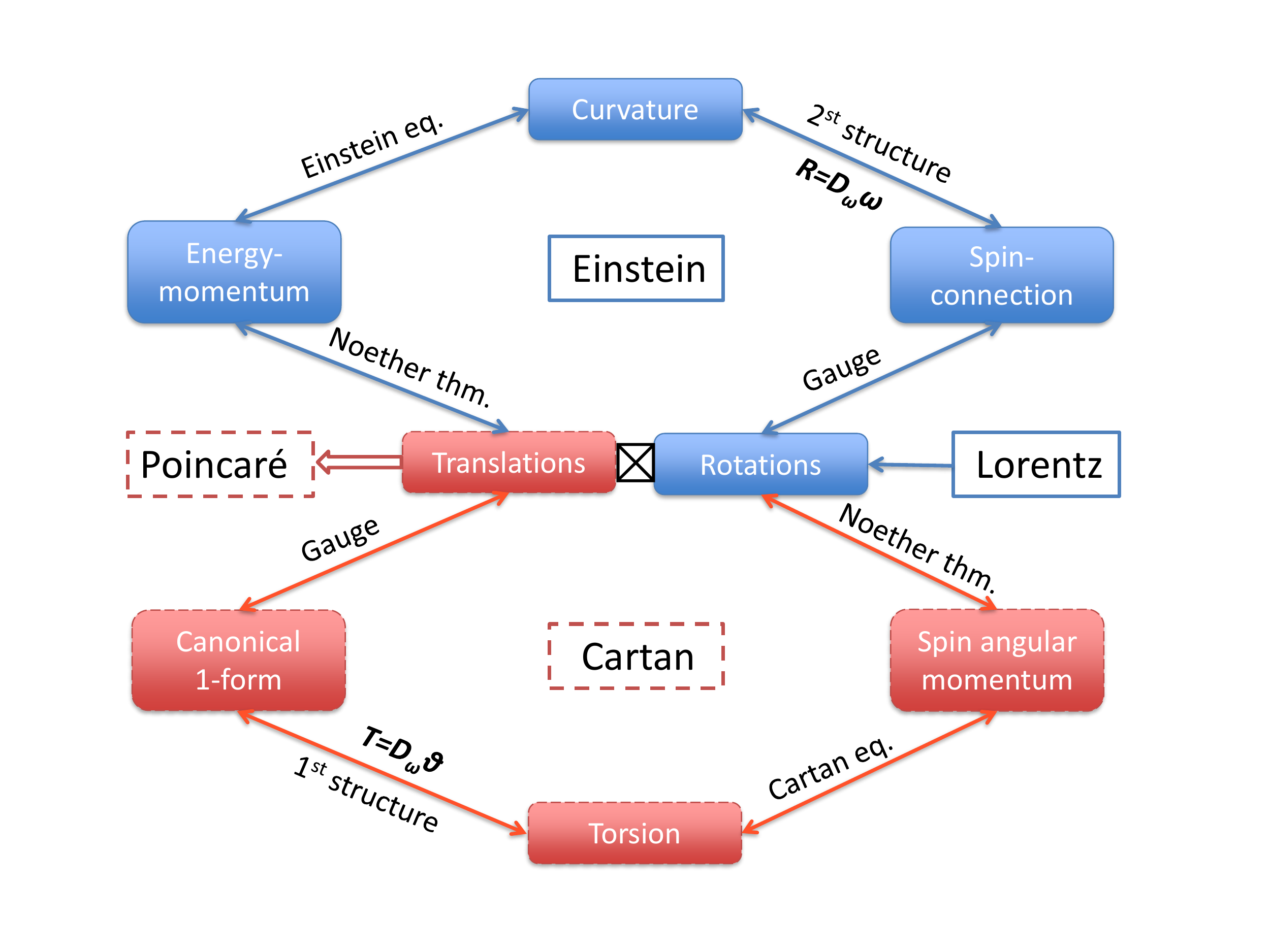}
	\caption{The logical framework of PGG. Poincar\'e group is the semidirect product of the translation group and the rotation group. A gauge theory based on the Poincar\'e group with non-vanishing gauge strengths --- curvature and torsion --- is very symmetrical on the structure. The restriction of torsion-free leads to that Einstein's theory is not the maximum gauge field theory of gravity.}
	\label{fig:logicalframe}
\end{figure}

General speaking, the crucial difference between PGG and GR-based theories, such as $f(R)$ gravity, is that the former removed the restriction of torsion-free on the connections.
However, the direct generalization from HE action will be back to Einstein's theory, when the spin tensor (\ref{eq:defSpinTensor}) of matter vanishes because of the algebraic Cartan equation, i.e. torsion can not propagate in this case.
This makes it necessary for us to generalize the action to obtain the propagating torsion in the vacuum.
The standard PGG Lagrangian with quadratic-strength reads: \cite{nester2017gravity}:
\begin{equation}
	\mathcal{L}_G\sim \Lambda+curvature+torsion^2+\frac{1}{\varrho}curvature^2,
\end{equation}
where $\Lambda$ is the cosmological constant, and $\varrho$ the parameter with certain dimension.
The additional quadratic terms come from the traces of field strengths in the internal space, which are at most second derivative if one regards the canonical $1$-forms (or the tetrads) and the spin-connections as the fundamental variables.
It seems that these terms will introduce the ghosts degrees of freedom when one considers the particle substance of gravity.
That would be something troublesome even for a simple modified gravity theory.
The existence of the ghosts is closely related to the fact that the modified equation of motion has orders of time-derivative higher than two, for example, scale factor $a$ will be fourth-order over time in the general quadratic curvature case in FLRW cosmology.
Due to Ostrogradsky's theorem \cite{ostrogradsky1850memoires}, a system is not (kinematically) stable if it is described by a non-degenerate higher time-derivative Lagrangian.
To avoid the ghosts, a bunch of scalar-tensor theories of gravity was introduced, such as the Horndeski theory and beyond \cite{kobayashi2019horndeski,langlois2016degenerate}.
Another way to evade Ostrogradsky's theorem is to break Lorentz invariance in the ultraviolet and include only high-order spatial derivative terms in the Lagrangian, while still keeping the time derivative terms to the second order.
This is exactly what Ho\v rava did recently \cite{hovrava2009quantum,wang2017hovrava}.
Besides, another recipe to treat the ghosts is not removing them from the action, while focusing on the higher-order instability in the equations of motion \cite{carroll2005cosmology}.
For the general second-order Lagrangian with propagating torsion, a systematical way to remove the ghosts and the tachyons was introduced in \cite{neville1980gravity,sezgin1980new} using spin projection operators.
The gauge fields $(\theta,\omega)$ can be decomposed irreducibly by $su(2)$ group into different spin modes by means of the weak-field approximation.
In addition to the graviton, three classes spin-$0^{\pm},1^{\pm},2^{\pm}$ modes of torsion were introduced.
\cite{sezgin1980new} studied the general quadratic Lagrangian with nine parameters and obtained the conditions on the parameters for not having ghosts and tachyons at massive and massless sectors, respectively.
In this work, to develop a good cosmology based on PGG, we will adopt their nine-parameter Lagrangian with the ghost- and tachyons-free conditions on parameters.
The Hamiltonian analysis of PGG for different modes can be found in \cite{yo1999hamiltonian,yo2002hamiltonian}, which tell us that the only safe modes of torsion are spin-$0^{\pm}$, corresponding to the scalar and pseudo-scalar components of torsion, respectively.

It's natural to apply the corresponding Poincar\'e gauge cosmology (PGC) on understanding the evolution of the Universe.
The last decade, a series of work \cite{minkevich2007regular,shie2008torsion,minkevich2009accelerating,chen2009cosmological,li2009torsion,baekler2011poincare,ao2010analytical,ao2012torsion,garkun2011numerical,minkevich2013some,ho2015general} (from both analytical and numerical approach) proved that it is possible to reproduce the late-time acceleration in PGC without ``dark energy''.
In Ref. \cite{minkevich2006analysis}, the authors discussed the early-time behaviors of the expanding solution of PGC with a scalar field, while in Ref. \cite{wang2009inflation}, power-law inflation was studied in an $R+R^2$ model of PGC without inflaton.
The motivation of our work is naturally raised, that is, to seek a model in PGC which can unify both the early- and the late-time evolutions without any priori hypothesis of inflaton and ``dark energy''.

This paper is organized as follows. In Sec. \ref{secII} we briefly summarize the fundamental notions of PGG, and introduce the corresponding cosmology by means of some assumptions, then we get the conservation law in this framework. In Sec. \ref{secIII} we derive the general cosmological equations of the nine-parameter Lagrangian with ghost- and tachyon-free conditions, then by choosing certain combinations of parameters, we obtain two analytical solutions which fit the early- and the late-time evolutions of the Universe, respectively; we also estimate the magnitudes of the model parameters according to the latest observations. We conclude and discuss our work in Sec. \ref{secIV}.

\section{Poincar\'e gauge cosmology}\label{secII}
In a Riemann-Cartan spacetime, the local invariance with respect to the internal symmetries of Poincar\'e group induces two sets of gauge potentials: the canonical $1$-forms $\theta$s corresponding to the translations and the spin-connections $\omega$s corresponding to the rotations.
The canonical $1$-form is the dual co-vector of the tetrad, i.e.
\begin{equation}
	\theta^a(e_b)=\delta^a_b.
\end{equation}
The tetrad can be decomposed with respect to an ordinary coordinate basis $\{\partial_\mu\}$, with a set of coefficients $e_a{}^\mu$,
\begin{equation}
	e_a=e_a{}^\mu\partial_\mu.
\end{equation}
Here, we use $a,b,c,d,e$ to represent the internal indices, and $\mu,\nu,\rho,\lambda,\sigma$ to represent the external (spacetime) indices.
In PGG, one chooses the orthonormal tetrad basis, which induces the following relation between the spacetime metric and the Minkowski metric:
\begin{equation}
	\eta_{ab}=e_a{}^\mu e_b{}^\nu g_{\mu\nu}.
\end{equation}
Therefore, two formalisms, the spacetime metric and the tetrad, are consistent.
On the other hand, according to the definitions of the spin-connection $\omega$ and the affine-connection $\Gamma$:
\begin{eqnarray}
    \nabla_{e_a}e_b&=&\omega^c{}_{ab}e_c,\\
	\nabla_\mu\partial_\nu&=&\Gamma^\rho{}_{\mu\nu}\partial_\rho,
\end{eqnarray}
one has the relation between the spin-connection and the affine-connection:
\begin{equation}
	\omega^b{}_{\lambda a}dx^\lambda=(e^b{}_\nu\partial_\lambda e_a{}^\nu+e^b{}_\nu\Gamma^\nu{}_{\lambda\mu} e_a{}^\mu)dx^\lambda,
\end{equation}
where $dx^\lambda$ is the dual co-vector of $\partial_\lambda$.
Thus the spin-connection $1$-form $\omega^b{}_{a}\equiv\omega^b{}_{\lambda a}dx^\lambda$ and the affine-connection $1$-form $\Gamma^\nu{}_{\mu}\equiv\Gamma^\nu{}_{\lambda\mu}dx^\lambda$ are consistent.

The gauge field strengths, torsion and curvature, which are $2$-forms, can be constructed by means of the gauge potentials in the following way:
\begin{eqnarray}
	T^c&=&d\theta^c+\omega^c{}_b\wedge\theta^b,\\
	R^d{}_c&=&d\omega^d{}_c+\omega^d{}_e\wedge\omega^e{}_c,
\end{eqnarray}
which are also called the Cartan $1$st- and $2$nd-structure, respectively.
On the other hand, the same geometric objects can be characterized by the metric $g$ and the affine-connection $\Gamma$ using the external indices, where the torsion tensor is defined as the anti-symmetric part of the affine-connection $\Gamma$:
\begin{equation}
T^{\rho}{}_{\mu\nu}\coloneqq\Gamma^{\rho}{}_{\mu\nu}-\Gamma^{\rho}{}_{\nu\mu}.
\end{equation}
If we omit the non-metricity part of the affine-connection, i.e. $Q\equiv-\nabla g=0$, metric, torsion and affine-connection fulfill the following relation:
\begin{equation}
\Gamma^{\rho}{}_{\mu\nu} = \{^{\rho}{}_{\mu\nu}\}+\frac{1}{2} \left( T^{\rho}{}_{\mu\nu}+T_{\mu}{}^{\rho}{}_{\nu}+T_{\nu}{}^{\rho}{}_{\mu} \right),
\label{eq:GammaToT}
\end{equation}
where $\{^{\rho}{}_{\mu\nu}\}$ is the Levi-Civita connection constructed by means of the metric directly.
Eq. (\ref{eq:GammaToT}) shows that instead of the pair of canonical $1$-form and spin-connection, the combination of metric and torsion can also represent the independent structures in Riemann-Cartan spacetime.
The definition of curvature by means of the affine-connection reads:
\begin{equation}\label{curvature_expr1}
R_{\mu\nu}{}{}^\rho{}_\sigma=2\partial_{[\mu}\Gamma^\rho{}_{\nu]\sigma}+2\Gamma^\rho{}_{[\mu|\lambda|}\Gamma^\lambda{}_{\nu]\sigma}.
\end{equation}
which implies torsion in it.
In the subsequent context, $\nabla$ represents the covariant derivative with respect to the affine-connection.
The torsion-free curvatures are labeled by a tilde, such as $\tilde{R}$.

In PGC, we assume that the cosmological principle is still valid, namely, our Universe is homogeneous and isotropic when viewed on a large enough scale.
This assumption alone determines the spacetime metric up to FLRW form:
\begin{equation}
ds^{2} = -dt^{2}+a^{2}(t)
\left[ d\mathbf{x}^{2}+K \frac{\left(\mathbf{x}\cdot d\mathbf{x} \right)^{2}}{1-K \mathbf{x}^{2}} \right],
\label{eq:FRWmetric}
\end{equation}
where $a(t)$ is the scale factor with $t$ being the cosmic time.
In this work, we consider the spatially flat case, $K=0$, which is consistent with the observations mentioned earlier in this article.
Meanwhile, according to \cite{tsamparlis1979cosmological}, it turns out that the only torsion tensors compatible with FLRW Universe are the time-like vector torsion and the time-like axial torsion \cite{capozziello2001geometric}, which can be expressed in terms of a scalar $h(t)$ and a pseudo-scalar $f(t)$,
\begin{equation}
T_{ij0} = a^2 h(t) \delta_{ij}, ~~ T_{ijk} = a^3 f(t) \epsilon_{ijk}, ~~ i,j,k=1,2,3.
\label{eq:TorsionScalar}
\end{equation}
In addition, we assume that the spin effect of matter doesn't appear on the cosmological scales.
Since the spin orientation for ordinary matter is random, the macroscopic space average of the spin vanishes.
The Weyssenhoff fluid was introduced into Einstein-Cartan theory \cite{obukhov1987weyssenhoff} to describe the spin source of torsion.
However, \cite{kuchowicz1976cosmology,boehmer2006homogeneous} indicate that the Weyssenhoff fluid is incompatible with the cosmological principle.
So we do not consider the Weyssenhoff fluid in our work.
Therefore, the spin tensor, defined as
\begin{equation}
S_{\rho}{}^{\mu\nu} \coloneqq \frac{1}{\sqrt{\vert g\vert}} \frac{\delta \left(\sqrt{\vert g\vert} \mathcal{L}_{m} \right)}{\delta T^{\rho}{}_{\mu\nu}},
\label{eq:defSpinTensor}
\end{equation}
vanishes \cite{lu2016cosmology}.
There is still no direct evidence to support the existence of ``dark energy''.
When we say ``dark energy'', it means that a priori hypothesis in our phenomenological models, such as the $\Lambda$CDM (cosmological constant with cold dark matter model) and the $w$CDM (``dark energy'' with parameterized EOS $w$ and cold dark matter model).
Here, we do not assume a priori the existence of ``dark energy'', while we just start from our current knowledge of the Universe: radiation, (baryonic and dark) matter, and the late-time acceleration.
From the perspectives of geometry, the power caused the acceleration is called the geometric ``dark energy'' in some references \cite{sahni2006reconstructing,copeland2006dynamics,nojiri2007introduction,frieman2008dark}.
Nonetheless, in our work, the energy-momentum tensor defined as
\begin{equation}
T_m^{\mu\nu} \coloneqq \frac{1}{\sqrt{\vert g\vert}} \frac{\delta \left(\sqrt{\vert g\vert} \mathcal{L}_{m} \right)}{\delta g_{\mu\nu}},
\label{eq:defEMTensor}
\end{equation}
contains radiation, baryonic and dark matter.

Noether's theorem in PGG implies the conservation laws for energy-momentum and angular momentum currents, respectively \cite{hehl1976general,blagojevic2013gauge}.
The vanishing of spin tensor $S_{\rho}{}^{\mu\nu}$ (\ref{eq:defSpinTensor}) leads to the conservation law of energy-momentum tensor as
\begin{equation}\label{conserved_em}
\nabla_\mu T_m^\mu{}_\nu+T_\mu T_m^\mu{}_\nu+T_\mu{}^\rho{}_\nu T_m^{\mu}{}_\rho=0,
\end{equation}
where $T_\mu\equiv T^\nu{}_{\mu\nu}$ is the trace of the torsion tensor.
It is obvious that (\ref{conserved_em}) is independent of the choice of the gravitational Lagrangian.
Its time-component equation on FLRW background reads
\begin{equation}\label{rho_point}
\dot{\rho}+3(1+w)H\rho=0,
\end{equation}
which is same as the one in SM.
The EOS $w=\tfrac{1}{3}, 0$ correspond to relativistic particles (photons and massless neutrinos, or say radiation, labeled by ``r''), and baryons (``b''), respectively.
Eq. (\ref{rho_point}) can be solved immediately:
\begin{equation}\label{evolutionofrho}
\rho_r=\rho_{r0}a^{-4},~~\rho_b=\rho_{b0}a^{-3},
\end{equation}
where the subscript ``$0$'' represents values at present, $a_0=1$.
In the $\Lambda$CDM model, the cosmological constant $\Lambda$ is regarded as the constant ``dark energy'' with $w=-1$.
To understand the phenomenon of the late-time acceleration, we solve Eq. (\ref{rho_point}) for $\Lambda$ also:
\begin{equation}
	\rho_\Lambda=const.
\end{equation}
\section{The solutions of PGC on FLRW background}\label{secIII}
\subsection{The general cosmological equations}
According to \cite{sezgin1980new,hayashi1980gravity}, we consider a nine-parameter gravitational Lagrangian $\mathcal{L}_G$, at most quadratic in torsion and curvature, which reads:
\begin{alignat}{2}\label{general_action}
I=&\int d^4x\sqrt{\vert g\vert}\big[\frac{1}{2\kappa}\mathcal{L}_G+\mathcal{L}_M\big],\nonumber\\
\mathcal{L}_G=&\alpha R+\mathcal{L}_T+\mathcal{L}_R,\nonumber\\
\mathcal{L}_T\equiv& a_1T_{\mu\nu\rho}T^{\mu\nu\rho}+a_2T_{\mu\nu\rho}T^{\nu\mu\rho}+a_3T_{\mu}T^{\mu},\nonumber\\
\mathcal{L}_R\equiv& b_1R_{\mu\nu\rho\sigma}R^{\mu\nu\rho\sigma}+b_2R_{\mu\nu\rho\sigma}R^{\rho\sigma\mu\nu}+b_3R_{\mu\nu}R^{\mu\nu}\nonumber\\
&+b_4R_{\mu\nu}R^{\nu\mu}+b_5R_{\mu\nu\rho\sigma}R^{\mu\rho\nu\sigma},
\end{alignat}
where $\kappa\equiv8\pi G$, and $\alpha$, $a_1\sim a_3$, $b_1\sim b_5$ are free Lagrangian parameters with appropriate units.
The $R^2$ term need not be included due to the use of the Chern-Gauss-Bonnet theorem \cite{chern1944simple}:
\begin{equation}
\int d^4x\sqrt{\vert g\vert}(R_{\mu\nu\rho\sigma}R^{\mu\nu\rho\sigma}-4R_{\mu\nu}R^{\mu\nu}+R^2)=0,
\end{equation}
for spacetime topologically equivalent to flat space.
For the pair of gauge fields ($\theta,\omega$) as the dynamical variables, the field equations are up to $2$nd-order.
However, the gauge fields ($\theta,\omega$) can be decomposed irreducibly by $su(2)$ group into different spin modes by means of the weak-field approximation.
In addition to the graviton, three classes spin-$0^{\pm},1^{\pm},2^{\pm}$ modes of torsion are introduced.
It is obvious that in such a general quadratic, the ghosts and tachyons are inevitable for certain modes.
Fortunately, the authors studied this Lagrangian in \cite{sezgin1980new} using the spin projection operators and obtained the conditions on parameters for not having ghosts and tachyons at massive and massless sectors, respectively.

According to \cite{sezgin1980new}, we summarize the ghost- and tachyon-free conditions on parameters for action (\ref{general_action}) in TABLE I:
\begin{table}[!htbp]
	\caption{The ghost- and tachyon-free conditions on parameters for six spin modes, respectively.}\label{ghost-free_conditions}
	\begin{tabularx}{\textwidth}{c|c}
		\hline\hline
		spin modes & conditions on parameters\\
		\hline
		\multirow{2}{*}{$2^-$}&$4 b_1+b_5<0$\\ &$\alpha +2a_1+a_2<0$\\
		\hline
		\multirow{2}{*}{$1^-$}&$4 b_1+2 b_3+b_5<0$\\ &$(\alpha +2a_1+a_2) (2a_1+a_2+a_3)(-2\alpha +2 a_1+a_2+3 a_3)<0$\\
		\hline
		\multirow{2}{*}{$0^-$}&$-2b_1+b_5>0$\\ &$\alpha-4a_1 +4 a_2>0$\\
		\hline
		\multirow{2}{*}{$2^+$}&$4 b_1+4b_2+b_3+b_4+2b_5>0$\\ &$\alpha  (2a_1+a_2) (\alpha +2a_1+a_2)<0$\\
		\hline
		\multirow{2}{*}{$1^+$}&$-4b_1+4 b_2-b_3+b_4<0$\\ &$(2a_1-a_2) (-\alpha +4a_1-4 a_2)(\alpha +2a_1+a_2)>0$\\
		\hline
		\multirow{2}{*}{$0^+$}&$b_1+b_2+b_3+b_4+b_5/2>0$\\ &$\alpha (2 a_1+a_2+3a_3)(-2 \alpha +2a_1+a_2+3a_3)>0$\\
		\hline\hline
	\end{tabularx}
\end{table}

For the massless sector, the ghost-free condition is just: $\alpha>0$.

The ghost- and tachyon-free conditions in Table \ref{ghost-free_conditions} are derived by means of the fundamental variables ($\theta,\omega$), but it's general in any case.
According to the previous context, it is convenient for us to treat $(g,T)$ as the fundamental variables to derive the field equations.
Varying the action (\ref{general_action}) with respect to $g_{\mu\nu}$ and $T^{\rho}{}_{\mu\nu}$, respectively, as well as considering (\ref{eq:defSpinTensor}) and (\ref{eq:defEMTensor}), one can get the modified Einstein and the modified Cartan field equations.
We do the calculations with the help of \texttt{xAct: Efficient tensor computer algebra for the Wolfram Language}\footnote{Authors: Jos\'e M. Mart\'in-Garc\'ia et. al. Homepage: \url{http://www.xact.es/}}.
We integrate our calculations in a Wolfram package \texttt{PGC: Symbolic computing package for Poincare Gauge Cosmology}\footnote{\texttt{PGC version 1.2.1}: \url{https://github.com/zhanghc0537/Poincare-Gauge-Cosmology}}, which is available on Github.
Since the field equations are too paper-consuming, we don't intend to copy them here, but feel free to download and install our package if you want to check the field equations and their components on FLRW background.
The \texttt{README} file will indicate you how to use it.
Here, we just sort out the cosmological equations on FLRW background which read:
\begin{eqnarray}
\tfrac{1}{3}\kappa\rho
&=&\alpha(H^2-h^2-f^2)+4A_0f^2+\tfrac{1}{2}A_1h^2\nonumber\\
&&+2B_0\big\{2(H-h)(\ddot{H}-\ddot{h})-(\dot{H}-\dot{h})^2+4H(\dot{H}-\dot{h})(H-h)-4f\dot{f}(H-h)\nonumber\\
&&+2\dot{H}(H-h)^2+[f^2-3h(H-h)][f^2-(H-h)(2H-h)]\big\}\nonumber\\
&&+2B_1[f^2(3H-2h)^2+2f\dot{f}(3H-2h)+\dot{f}^2]+4B_2f(fH+\dot{f})(H-h),\label{general_ceq_1}\\
-\kappa w\rho &=&\alpha(2\dot{H}+3H^2+3h^2-f^2)+4A_0f^2-\tfrac{3}{2}A_1h^2\nonumber\\
&&+2B_0\big[2(H^{(3)}-h^{(3)})+12H(\ddot{H}-\ddot{h})+2(H-h)\ddot{h}-4f\ddot{f}+9\dot{H}^2-10\dot{H}\dot{h}\nonumber\\
&&+18H^2\dot{H}+18hH\dot{H}-14h^2\dot{H}+\dot{h}^2-6H^2\dot{h}-28hH\dot{h}+12h^2\dot{h}-4f^2\dot{h}-4\dot{f}^2\nonumber\\
&&-8fH\dot{f}-12fh\dot{f}+(f^2+3h^2+9hH)(-f^2+h^2-3hH+2H^2)\big]\nonumber\\
&&+2B_1\big[4f\ddot{f}+12f^2\dot{H}-8f^2\dot{h}+3\dot{f}^2+30fH\dot{f}-12fh\dot{f}+f^2(27H^2-12hH-4h^2)\big]\nonumber\\
&&+4B_2\big[f\ddot{f}+f^2\dot{H}+\dot{f}^2+4fH\dot{f}+fh\dot{f}+2f^2H^2+f^2hH\big],\label{general_ceq_2}\\
0&=&\alpha h-\tfrac{1}{2}A_1h\nonumber\\
&&+2B_0\big[\ddot{H}-\ddot{h}+4H\dot{H}-h\dot{H}-3H\dot{h}-2f\dot{f}+4hH^2-6h^2H+2h^3-2f^2h\big]\nonumber\\
&&+4B_1f(\dot{f}+3fH-2fh)+2B_2f(\dot{f}+fH),\label{general_ceq_3}\\
0&=&f\big\{\alpha-4A_0+4B_0\big[\dot{H}-\dot{h}+2H^2-3hH+h^2-f^2\big]\nonumber\\
&&-2B_1\big[(3\dot{H}-2\dot{h})+2h(3H-2h)\big]-2B_2\big[(\dot{H}-\dot{h})+H(H-h)\big]\big\}\nonumber\\
&&-2B_1(3H\dot{f}+\ddot{f}),\label{general_ceq_4}
\end{eqnarray}
with the combinations of parameters
\begin{alignat}{2}
A_0&\equiv a_1-a_2,~~A_1\equiv 2a_1+a_2+3a_3,\nonumber\\
B_0&\equiv b_1+b_2+b_3+b_4+\frac{1}{2}b_5,\nonumber\\
B_1&\equiv b_1-\frac{1}{2}b_5,~~B_2\equiv 4b_2+b_3+b_4+b_5.
\end{alignat}
(\ref{general_ceq_1}) is the time-component of the modified Einstein field equation, and (\ref{general_ceq_2}) is the trace of its space-component.
While (\ref{general_ceq_3},\ref{general_ceq_4}) are the non-vanishing components of the modified Cartan field equation corresponding to the evolution of $h$ and $f$, respectively.
The degeneracy among these Lagrangian parameters on the background makes the inequalities in Table \ref{ghost-free_conditions} can not be solved completely.
However, it is obvious that ghost- and tachyon-free spin-$0^{\pm}$ ``particles'' require:
\begin{equation}\label{ghost-free-spin0}
	B_0>0,~~B_1<0,~~\alpha-4A_0>0,~~\alpha A_1(A_1-\alpha)>0.
\end{equation}
The above constraints for not having ghosts and tachyons follow from the requirement of having real mass and positive-definite residue matrix at the pole \cite{sezgin1980new}.
If one changes the inequality to an equal sign, the corresponding term may be lost the mass or momentum, therefore, the ``particle'' becomes massless or non-dynamic.
We will consider this case by forcing the spin-$0^{-}$ mode non-dynamic and the spin-$0^{+}$ mode massless, such that the system can be solved analytically.

(\ref{general_ceq_1}) and (\ref{general_ceq_2}) correspond to the generalized Friedmann equations.
For more discussions on the cosmological aspects of these field equations, please see \cite{minkevich2007regular,shie2008torsion,minkevich2009accelerating,chen2009cosmological,li2009torsion,baekler2011poincare,ao2010analytical,ao2012torsion,garkun2011numerical,minkevich2013some,ho2015general}.
\subsection{The analytical solutions}
Two independent Friedmann-like solutions can be obtained from (\ref{general_ceq_1}) $\sim$ (\ref{general_ceq_4}), if we just set:
\begin{eqnarray}\label{constraints_parameters}
A_1=B_1=B_2=0,~4A_0=1,
\end{eqnarray}
which corresponding to the spin-$0^{-}$ mode non-dynamic and the spin-$0^{+}$ mode massless case.
Then the cosmological solutions read:
\begin{alignat}{2}
H^2&=\frac{\kappa\rho}{3\alpha}\frac{[3\alpha^2+2(1-3w)B_0\kappa\rho][3\alpha^2+\frac{(1-3w)^2}{2}B_0\kappa\rho]}{[3\alpha^2-(1+3w)(1-3w)B_0\kappa\rho]^2}\nonumber\\
&=
\begin{cases}
\frac{\kappa\rho_r}{3\alpha}& w=\frac{1}{3}\\
\frac{\kappa\rho_b}{3\alpha}\frac{(3\alpha^2+\frac{1}{2}B_0\kappa\rho_b)(3\alpha^2+2B_0\kappa\rho_b)}{(3\alpha^2-B_0\kappa\rho_b)^2}& w=0\\
\frac{\kappa\rho_\Lambda}{3\alpha}& w=-1
\end{cases},\label{h_BGeq_H2}\\
h^2&=\frac{3B_0^2(1+w)^2(1-3w)^2(\kappa\rho)^3[3\alpha^2+\frac{(1-3w)^2}{2}B_0\kappa\rho]}{\alpha[3\alpha^2+2(1-3w)B_0\kappa\rho][3\alpha^2-(1+3w)(1-3w)B_0\kappa\rho]^2}\nonumber\\
&=
\begin{cases}
0& w=\frac{1}{3}\\
\frac{3B_0^2(\kappa\rho_b)^3(3\alpha^2+\frac{1}{2}B_0\kappa\rho_b)}{(3\alpha^2+2B_0\kappa\rho_b)(3\alpha^2-B_0\kappa\rho_b)^2}& w=0\\
0& w=-1
\end{cases},\label{h_BGeq_h2}\\
f&=0\label{h_BGeq_f},
\end{alignat}
or
\begin{alignat}{2}
H^2&=\frac{1}{3}\big[\kappa\rho+\frac{3(\alpha-1)^2}{8B_0}\big],\label{f_BGeq_H2}\\
f^2&=\frac{1-3w}{6}\kappa\rho+\frac{\alpha(\alpha-1)}{4B_0},\label{f_BGeq_f2}\\
h&=0\label{f_BGeq_h}.
\end{alignat}
It's obvious that if $\alpha\rightarrow1$ and $B_0\rightarrow0$, the first solution ($h$-determined) reduces to the SM.
However, this solution doesn't contain a natural ``dark energy'' ($\rho_\Lambda$ arises only when we set $w=-1$ artificially, which means the ``dark energy'' is introduced in the same way as SM), while the second solution ($f$-determined) implies a constant $\frac{3(\alpha-1)^2}{8B_0}$ if $\alpha\neq1$ and $B_0\neq0$.
Therefore, $\alpha$ must be greater (according to (\ref{ghost-free-spin0}) and (\ref{constraints_parameters})) but very close to $1$, and $B_0$ can be estimated by the value of ``dark energy'' density parameterized in the $\Lambda$CDM.
$\alpha$ and $B_0$ represent the weights of Ricci scalar and the bulk effect of the quadratic curvatures, respectively.
And this constant $\frac{3(\alpha-1)^2}{8B_0}$ can be regarded as the coupling strength between the curvature field and its quadratic terms.
On the other hand, the first solution is more consistent with the evolution in the early-time.
Since during the reheating (at any given temperature $T$), the energy density of non-relativistic particles in thermal equilibrium is exponentially suppressed with respect to that of relativistic particles, which means that during the reheating:
\begin{equation}\label{reheating}
w=\frac{1}{3},~~\rho=\rho_r=\frac{\pi^2}{30}g_{*}T^4.
\end{equation}
The first solution shows $h=0, f=0$ naturally when $w=\tfrac{1}{3}$.
Substituting $h=0, f=0$ into the general field equations and considering (\ref{constraints_parameters}), (\ref{general_ceq_1}) reduces to
\begin{equation}
	4B_0H\ddot{H}-2B_0\dot{H}^2+12B_0H^2\dot{H}+\alpha H^2=\frac{\kappa\rho}{3},
\end{equation}
which is consistent with the Starobinsky cosmology.
And the Starobinsky inflation arises if $\rho$ vanishes.
$\tfrac{3\alpha}{B_0}$ is equivalent to $6M^2$ in the standard Starobinsky inflation:
\begin{equation}\label{Starobinsky_action}
\mathcal{L}_{\tilde{G}}=\tilde{R}+\frac{1}{6M^2}\tilde{R}^2,
\end{equation}
where $M$ is the scalaron rest-mass, determined from the normalization of the primordial scalar spectrum \cite{starobinsky2007disappearing}.

\subsection{The numerical magnitudes of parameters}
The combined WMAP3-SDSS measurements \cite{tegmark2006cosmological} constrain $M=2.8\times10^{-6}(N/50)^{-1}M_{Pl}$ where $M_{Pl}\equiv1/\sqrt{G}$ is the Planck mass \footnote{The natural units are used in this subsection, where $\hbar=c=1$}, $N$ the number of e-folds between the first Hubble radius crossing of the present inverse comoving scale $0.05 Mpc^{-1}$ and the end of inflation.
The model predictions for the slope of the primordial spectrum of scalar perturbations $n_s$ is $n_s-1=-2N^{-1}$.
\texttt{Planck 2018} \cite{akrami2018planck} temperature, polarization, and lensing data determine the spectral index of scalar perturbations to be $n_s=0.9649\pm0.0042$ at $68\%$ CL.
\texttt{Planck 2018} also constrain the inferred late-Universe parameters of the $\Lambda$CDM are: Hubble constant $H_0=(67.4\pm0.5)km~sec^{-1}Mpc^{-1}$; matter density parameter $\Omega_m=0.315\pm0.007$.
Ignoring the spatial curvature $\Omega_K$ and radiation $\Omega_r$, the ``dark energy'' $\kappa\rho_\Lambda\simeq(1-\Omega_m)\kappa\rho_{crit}=(1-\Omega_m)(3H^2_0)\simeq4.25\times10^{-84}GeV^{2}$.
Combining the above data, we get the constraints (omitting the error bars):
\begin{alignat}{2}
	\frac{3\alpha}{B_0}&=6M^2\simeq5.22\times 10^{27}GeV^2,\nonumber\\
	\frac{3(\alpha-1)^2}{8B_0}&=\kappa\rho_\Lambda\simeq 4.25\times10^{-84}GeV^{2},
\end{alignat}
which derive
\begin{alignat}{2}
	\alpha-1&\simeq 8.07\times 10^{-56},\nonumber\\
	B_0&\simeq 5.76\times 10^{-28}GeV^{-2}.
\end{alignat}
According to these estimations, $B_0\kappa\rho_b\ll\alpha^2$, thus $H^2\simeq\frac{\kappa\rho_b}{3\alpha}$, $h^2\simeq\frac{B_0^2(\kappa\rho_b)^3}{3\alpha^4}$ when $w=0$ in (\ref{h_BGeq_H2},\ref{h_BGeq_h2}).
$\alpha$ is $-54$ order of magnitudes close to $1$, which hit our previous guess.
(\ref{h_BGeq_h2},\ref{f_BGeq_f2}) show that the evolution of the Universe on the background is influenced by the spacetime torsion besides the effects from curvature.
Conversely, such an evolution of the Universe proves indirectly the existence of torsion.
In Ref. \cite{minkevich2012limiting,hehl2013poincare,minkevich2019gravitational}, authors discussed the practicability of the direct measurement of the spacetime torsion, which will help us further study the evolutionary mechanism of the Universe in PGC.
\section{Conclusion}\label{secIV}
In this work, we start from the construction of Poincar\'e gauge cosmology based on certain assumptions and derive the conservation law (\ref{rho_point}) in this framework, which is still the usual one in SM.
Then by varying the general nine-parameter PGC Lagrangian (\ref{general_action}), we get the general cosmological equations (\ref{general_ceq_1}$\sim$\ref{general_ceq_4}).
To remove the ghosts and tachyons in this Lagrangian, we summarized the ghost- and tachyon-free conditions on parameters in TABLE I according to \cite{sezgin1980new}.
Two Friedmann-like analytical solutions are obtained when the constraints for Lagrangian parameters (\ref{constraints_parameters}) are imposed.
The $h$-determined solution is consistent with the Starobinsky cosmology in the early-time, and the $f$-determined one contains naturally a constant $3(\alpha-1)^2/8B_0$, which can be regarded as the ``dark energy'' density.
These two solutions are consistent because they are obtained from the same gravitational Lagrangian.
According to the latest observations, and the behaviors in the early- and late-time respectively, we estimate the magnitudes of parameters: $\alpha-1\simeq 8.07\times10^{-56}$, $B_0\simeq5.76\times10^{-28}GeV^{-2}$ in the natural units.
(\ref{general_ceq_4}) shows that $B_1=0$ leads to the second-order term of $f$ vanished, i.e. (\ref{general_ceq_4}) is just a constraint, which is the key reason to get these two solutions.
Loosing the constraint on $B_1$ to an infinitesimal level, such that the $h$-determined solution is an asymptotic solution in the early-time and the $f$-determined solution is an asymptotic solution in the late-time, which will be the methodology to pursue a continuous model of PGC, where the inflation and late-time acceleration can be unified smoothly.
For more general case (without the constraint (\ref{constraints_parameters}) on parameters) and the inflation in PGC, please see our recent preprint \cite{zhang2019inflation}.
We expect that further analysis PGC can be consistent with the recent evidence of dynamic ``dark energy'' \cite{zhao2017dynamical} from the perspective of geometry and gauge field theory.

\begin{acknowledgments}
We thank Prof. Abhay Ashtekar for helpful comments.
Lixin Xu is supported in part by National Natural Science Foundation of China under Grant No. 11675032.
Hongchao Zhang is supported by the program of China Scholarships Council No. 201706060084.
Our calculations are accomplished with the help of a series of Wolfram packages under the name \texttt{xAct}, written by Jos\'e M. Mart\'in-Garc\'ia et. al.
\end{acknowledgments}

\bibliographystyle{JHEP}
\bibliography{PGGC_jcap}

\providecommand{\href}[2]{#2}\begingroup\raggedright\begin{thebibliography}{10}

\bibitem{aghanim2018planck}
N.~Aghanim, Y.~Akrami, M.~Ashdown, J.~Aumont, C.~Baccigalupi, M.~Ballardini
  et~al., \emph{Planck 2018 results. vi. cosmological parameters}, {\emph{arXiv
  preprint arXiv:1807.06209} (2018) }.

\bibitem{copeland2006dynamics}
E.~J. Copeland, M.~Sami and S.~Tsujikawa, \emph{Dynamics of dark energy},
  {\emph{International Journal of Modern Physics D} {\bfseries 15} (2006)
  1753}.

\bibitem{akrami2018planck}
Y.~Akrami, F.~Arroja, M.~Ashdown, J.~Aumont, C.~Baccigalupi, M.~Ballardini
  et~al., \emph{Planck 2018 results. x. constraints on inflation}, {\emph{arXiv
  preprint arXiv:1807.06211} (2018) }.

\bibitem{starobinsky1980new}
A.~A. Starobinsky, \emph{A new type of isotropic cosmological models without
  singularity}, {\emph{Physics Letters B} {\bfseries 91} (1980) 99}.

\bibitem{vilenkin1985classical}
A.~Vilenkin, \emph{Classical and quantum cosmology of the starobinsky
  inflationary model}, {\emph{Physical Review D} {\bfseries 32} (1985) 2511}.

\bibitem{mijic1986r}
M.~B. Miji{\'c}, M.~S. Morris and W.-M. Suen, \emph{The r 2 cosmology:
  Inflation without a phase transition}, {\emph{Physical Review D} {\bfseries
  34} (1986) 2934}.

\bibitem{castellanos2018higher}
A.~R.~R. Castellanos, F.~Sobreira, I.~L. Shapiro and A.~A. Starobinsky,
  \emph{On higher derivative corrections to the r+ r2 inflationary model},
  {\emph{Journal of Cosmology and Astroparticle Physics} {\bfseries 2018}
  (2018) 007}.

\bibitem{nojiri2003modified}
S.~Nojiri and S.~D. Odintsov, \emph{Modified gravity with negative and positive
  powers of curvature: Unification of inflation and cosmic acceleration},
  {\emph{physical Review D} {\bfseries 68} (2003) 123512}.

\bibitem{nojiri2007unifying}
S.~Nojiri and S.~D. Odintsov, \emph{Unifying inflation with $\lambda$cdm epoch
  in modified f (r) gravity consistent with solar system tests}, {\emph{Physics
  Letters B} {\bfseries 657} (2007) 238}.

\bibitem{cognola2008class}
G.~Cognola, E.~Elizalde, S.~Nojiri, S.~D. Odintsov, L.~Sebastiani and
  S.~Zerbini, \emph{Class of viable modified f (r) gravities describing
  inflation and the onset of accelerated expansion}, {\emph{Physical Review D}
  {\bfseries 77} (2008) 046009}.

\bibitem{nojiri2008modified}
S.~Nojiri and S.~D. Odintsov, \emph{Modified f (r) gravity unifying r m
  inflation with the $\lambda$ cdm epoch}, {\emph{Physical Review D} {\bfseries
  77} (2008) 026007}.

\bibitem{elizalde2011nonsingular}
E.~Elizalde, S.~Nojiri, S.~D. Odintsov, L.~Sebastiani and S.~Zerbini,
  \emph{Nonsingular exponential gravity: a simple theory for early-and
  late-time accelerated expansion}, {\emph{Physical Review D} {\bfseries 83}
  (2011) 086006}.

\bibitem{nojiri2011unified}
S.~Nojiri and S.~D. Odintsov, \emph{Unified cosmic history in modified gravity:
  from f (r) theory to lorentz non-invariant models}, {\emph{Physics Reports}
  {\bfseries 505} (2011) 59}.

\bibitem{elizalde2017beyond}
E.~Elizalde, S.~Odintsov, L.~Sebastiani and R.~Myrzakulov,
  \emph{Beyond-one-loop quantum gravity action yielding both inflation and
  late-time acceleration}, {\emph{Nuclear Physics B} {\bfseries 921} (2017)
  411}.

\bibitem{odintsov2017unification}
S.~Odintsov, V.~Oikonomou and L.~Sebastiani, \emph{Unification of constant-roll
  inflation and dark energy with logarithmic r2-corrected and exponential f (r)
  gravity}, {\emph{Nuclear Physics B} {\bfseries 923} (2017) 608}.

\bibitem{hehl1976general}
F.~W. Hehl, P.~Von~der Heyde, G.~D. Kerlick and J.~M. Nester, \emph{General
  relativity with spin and torsion: Foundations and prospects}, {\emph{Reviews
  of Modern Physics} {\bfseries 48} (1976) 393}.

\bibitem{blagojevic2013gauge}
M.~Blagojevi{\'c}, F.~W. Hehl and T.~Kibble, \emph{Gauge theories of
  gravitation: a reader with commentaries}. World Scientific, 2013.

\bibitem{nester2017gravity}
J.~M. Nester and C.-M. Chen, \emph{Gravity: A gauge theory perspective},  in
  \emph{Everything about Gravity: Proceedings of the Second LeCosPA
  International Symposium}, pp.~8--18, World Scientific, 2017.

\bibitem{ostrogradsky1850memoires}
M.~Ostrogradsky, \emph{M{\'e}moires sur les {\'e}quations diff{\'e}rentielles,
  relatives au probl{\`e}me des isop{\'e}rim{\`e}tres}, {\emph{Mem. Acad. St.
  Petersbourg} {\bfseries 6} (1850) 385}.

\bibitem{kobayashi2019horndeski}
T.~Kobayashi, \emph{Horndeski theory and beyond: a review}, {\emph{arXiv
  preprint arXiv:1901.07183} (2019) }.

\bibitem{langlois2016degenerate}
D.~Langlois and K.~Noui, \emph{Degenerate higher derivative theories beyond
  horndeski: evading the ostrogradski instability}, {\emph{Journal of Cosmology
  and Astroparticle Physics} {\bfseries 2016} (2016) 034}.

\bibitem{hovrava2009quantum}
P.~Ho{\v{r}}ava, \emph{Quantum gravity at a lifshitz point}, {\emph{Physical
  Review D} {\bfseries 79} (2009) 084008}.

\bibitem{wang2017hovrava}
A.~Wang, \emph{Ho{\v{r}}ava gravity at a lifshitz point: a progress report},
  {\emph{International Journal of Modern Physics D} {\bfseries 26} (2017)
  1730014}.

\bibitem{carroll2005cosmology}
S.~M. Carroll, A.~De~Felice, V.~Duvvuri, D.~A. Easson, M.~Trodden and M.~S.
  Turner, \emph{Cosmology of generalized modified gravity models},
  {\emph{Physical Review D} {\bfseries 71} (2005) 063513}.

\bibitem{neville1980gravity}
D.~E. Neville, \emph{Gravity theories with propagating torsion},
  {\emph{Physical Review D} {\bfseries 21} (1980) 867}.

\bibitem{sezgin1980new}
E.~Sezgin and P.~van Nieuwenhuizen, \emph{New ghost-free gravity lagrangians
  with propagating torsion}, {\emph{Physical Review D} {\bfseries 21} (1980)
  3269}.

\bibitem{yo1999hamiltonian}
H.-J. Yo and J.~M. Nester, \emph{Hamiltonian analysis of poincar{\'e} gauge
  theory scalar modes}, {\emph{International Journal of Modern Physics D}
  {\bfseries 8} (1999) 459}.

\bibitem{yo2002hamiltonian}
H.-J. Yo and J.~M. Nester, \emph{Hamiltonian analysis of poincar{\'e} gauge
  theory: higher spin modes}, {\emph{International Journal of Modern Physics D}
  {\bfseries 11} (2002) 747}.

\bibitem{minkevich2007regular}
A.~Minkevich, A.~Garkun and V.~Kudin, \emph{Regular accelerating universe
  without dark energy in poincar{\'e} gauge theory of gravity},
  {\emph{Classical and Quantum Gravity} {\bfseries 24} (2007) 5835}.

\bibitem{shie2008torsion}
K.-F. Shie, J.~M. Nester and H.-J. Yo, \emph{Torsion cosmology and the
  accelerating universe}, {\emph{Physical Review D} {\bfseries 78} (2008)
  023522}.

\bibitem{minkevich2009accelerating}
A.~V. Minkevich, \emph{Accelerating universe with spacetime torsion but without
  dark matter and dark energy}, {\emph{Physics Letters B} {\bfseries 678}
  (2009) 423}.

\bibitem{chen2009cosmological}
H.~Chen, F.-H. Ho, J.~M. Nester, C.-H. Wang and H.-J. Yo, \emph{Cosmological
  dynamics with propagating lorentz connection modes of spin zero},
  {\emph{Journal of Cosmology and Astroparticle Physics} {\bfseries 2009}
  (2009) 027}.

\bibitem{li2009torsion}
X.-z. Li, C.-b. Sun and P.~Xi, \emph{Torsion cosmological dynamics},
  {\emph{Physical Review D} {\bfseries 79} (2009) 027301}.

\bibitem{baekler2011poincare}
P.~Baekler, F.~W. Hehl and J.~M. Nester, \emph{Poincar{\'e} gauge theory of
  gravity: Friedman cosmology with even and odd parity modes: Analytic part},
  {\emph{Physical Review D} {\bfseries 83} (2011) 024001}.

\bibitem{ao2010analytical}
X.-c. Ao, X.-z. Li and P.~Xi, \emph{Analytical approach of late-time evolution
  in a torsion cosmology}, {\emph{Physics Letters B} {\bfseries 694} (2010)
  186}.

\bibitem{ao2012torsion}
X.-C. Ao and X.-Z. Li, \emph{Torsion cosmology of poincare gauge theory and the
  constraints of its parameters via sneia data}, {\emph{Journal of Cosmology
  and Astroparticle Physics} {\bfseries 2012} (2012) 003}.

\bibitem{garkun2011numerical}
A.~Garkun, V.~Kudin, A.~Minkevich and Y.~G. Vasilevsky, \emph{Numerical
  analysis of cosmological models for accelerating universe in poincar{\'e}
  gauge theory of gravity}, {\emph{arXiv preprint arXiv:1107.1566} (2011) }.

\bibitem{minkevich2013some}
A.~Minkevich, A.~Garkun and V.~Kudin, \emph{On some physical aspects of
  isotropic cosmology in riemann-cartan spacetime}, {\emph{Journal of Cosmology
  and Astroparticle Physics} {\bfseries 2013} (2013) 040}.

\bibitem{ho2015general}
F.-H. Ho, H.~Chen, J.~M. Nester and H.-J. Yo, \emph{General
  poincar$\backslash$'e gauge theory cosmology}, {\emph{arXiv preprint
  arXiv:1512.01202} (2015) }.

\bibitem{minkevich2006analysis}
A.~Minkevich and A.~Garkun, \emph{Analysis of inflationary cosmological models
  in gauge theories of gravitation}, {\emph{Classical and Quantum Gravity}
  {\bfseries 23} (2006) 4237}.

\bibitem{wang2009inflation}
C.-H. Wang and Y.-H. Wu, \emph{Inflation in r+ r2 gravity with torsion},
  {\emph{Classical and Quantum Gravity} {\bfseries 26} (2009) 045016}.

\bibitem{tsamparlis1979cosmological}
M.~Tsamparlis, \emph{Cosmological principle and torsion}, {\emph{Physics
  Letters A} {\bfseries 75} (1979) 27}.

\bibitem{capozziello2001geometric}
S.~Capozziello, G.~Lambiase and C.~Stornaioloi, \emph{Geometric classification
  of the torsion tensor of space-time}, {\emph{Annalen der Physik} {\bfseries
  10} (2001) 713}.

\bibitem{obukhov1987weyssenhoff}
Y.~N. Obukhov and V.~Korotky, \emph{The weyssenhoff fluid in einstein-cartan
  theory}, {\emph{Classical and Quantum Gravity} {\bfseries 4} (1987) 1633}.

\bibitem{kuchowicz1976cosmology}
B.~Kuchowicz, \emph{Cosmology with spin and torsion. part ii. spatially
  homogeneous aligned spin models with the weyssenhoff fluid.}, {\emph{Acta
  Cosmologica} {\bfseries 4} (1976) 67}.

\bibitem{boehmer2006homogeneous}
C.~G. Boehmer and P.~Bronowski, \emph{The homogeneous and isotropic weyssenhoff
  fluid}, {\emph{arXiv preprint gr-qc/0601089} (2006) }.

\bibitem{lu2016cosmology}
J.~Lu and G.~Chee, \emph{Cosmology in poincar{\'e} gauge gravity with a
  pseudoscalar torsion}, {\emph{Journal of High Energy Physics} {\bfseries 5}
  (2016) }.

\bibitem{sahni2006reconstructing}
V.~Sahni and A.~Starobinsky, \emph{Reconstructing dark energy},
  {\emph{International Journal of Modern Physics D} {\bfseries 15} (2006)
  2105}.

\bibitem{nojiri2007introduction}
S.~Nojiri and S.~D. Odintsov, \emph{Introduction to modified gravity and
  gravitational alternative for dark energy}, {\emph{International Journal of
  Geometric Methods in Modern Physics} {\bfseries 4} (2007) 115}.

\bibitem{frieman2008dark}
J.~A. Frieman, M.~S. Turner and D.~Huterer, \emph{Dark energy and the
  accelerating universe}, {\emph{Annu. Rev. Astron. Astrophys.} {\bfseries 46}
  (2008) 385}.

\bibitem{hayashi1980gravity}
K.~Hayashi and T.~Shirafuji, \emph{Gravity from poincar{\'e} gauge theory of
  the fundamental particles. i: General formulation}, {\emph{Progress of
  Theoretical Physics} {\bfseries 64} (1980) 866}.

\bibitem{chern1944simple}
S.-s. Chern, \emph{A simple intrinsic proof of the gauss-bonnet formula for
  closed riemannian manifolds}, {\emph{Annals of mathematics} (1944) 747}.

\bibitem{starobinsky2007disappearing}
A.~A. Starobinsky, \emph{Disappearing cosmological constant in f (r) gravity},
  {\emph{JETP Letters} {\bfseries 86} (2007) 157}.

\bibitem{tegmark2006cosmological}
M.~Tegmark, D.~J. Eisenstein, M.~A. Strauss, D.~H. Weinberg, M.~R. Blanton,
  J.~A. Frieman et~al., \emph{Cosmological constraints from the sdss luminous
  red galaxies}, {\emph{Physical Review D} {\bfseries 74} (2006) 123507}.

\bibitem{minkevich2012limiting}
A.~Minkevich, \emph{Limiting energy density and a regular accelerating universe
  in riemann-cartan spacetime}, {\emph{JETP letters} {\bfseries 94} (2012)
  831}.

\bibitem{hehl2013poincare}
F.~W. Hehl, Y.~N. Obukhov and D.~Puetzfeld, \emph{On poincar{\'e} gauge theory
  of gravity, its equations of motion, and gravity probe b}, {\emph{Physics
  Letters A} {\bfseries 377} (2013) 1775}.

\bibitem{minkevich2019gravitational}
A.~Minkevich, \emph{About gravitational interaction in astrophysics in
  riemann--cartan space-time}, {\emph{Classical and Quantum Gravity} {\bfseries
  36} (2019) 055003}.

\bibitem{zhang2019inflation}
H.~Zhang and L.~Xu, \emph{Inflation in the general poincar$\backslash$'e gauge
  cosmology}, {\emph{arXiv preprint arXiv:1906.04340} (2019) }.

\bibitem{zhao2017dynamical}
G.-B. Zhao, M.~Raveri, L.~Pogosian, Y.~Wang, R.~G. Crittenden, W.~J. Handley
  et~al., \emph{Dynamical dark energy in light of the latest observations},
  {\emph{Nature Astronomy} {\bfseries 1} (2017) 627}.

\end{thebibliography}\endgroup

\end{document}